\documentclass{webofc}
\usepackage[varg]{txfonts}   
\begin{document}
\title{Probing BFKL dynamics at hadronic colliders in  jet gap jet events}
%
%

\author{\firstname{Christophe} \lastname{Royon}\inst{1}\fnsep\thanks{\email{christophe.royon@ku.edu}}
\firstname{Federico} \lastname{Deganutti}\inst{1}\fnsep\thanks{\email{fedeganutti@gmail.com} }}

\institute{Department of Physics and Astronomy, The University of Kansas, Lawrence, USA 
          }

\abstract{%
In this report, we give the Balitsky Fadin Kuraev Lipton formalism for jet gap jet events at hadronic colliders. We also
discuss the case where in addition at least one proton is intact in the final state in diffractive events.
}
\maketitle
\section{Looking for BFKL resummation effects at hadronic colliders: forward jets and Mueller-Navelet jets}

The experimental search for resummation effects at low $x$ where $x$ is the momentum fraction of the interacting parton inside 
the proton, as predicted by the Balitsky Fadin Kuraev Lipatov (BFKL)~\cite{bfkl} equation has been an important topic of research
at recent colliders
The starting point to look for BFKL resummation effects at colliders was to study forward jet production
at HERA, the $ep$ collider located at DESY, Hamburg. When $Q$, the transferred energy between the electron and the interacting quark, is of the same order of the forward jet transverse momentum, and when the interval in rapidity between 
the forward jet and the scattered electron is large, the cross section predicted by usual Quantum Chromodynamics (QCD)
as given by the Dokhitzer Gribov Lipatov Altarelli Parisi (DGLAP)~\cite{dglap} evolution equation is small because of the
$k_T$-ordering of the different emitted gluons on the ladder. On the contrary, the BFKL dynamics predicts a non-negligible
cross section. As an example, the H1 collaboration measured the triple differential forward jet cross section as a function
of $Q^2$, jet $p_T$ and jet fractional energy~\cite{h1fwd}. The BFKL NLL formalism leads to a good description of data whereas the NLO QCD calculation has difficulties to describe the forward jet cross section~\cite{usfwd}. However
this measurement was not completely conclusive concerning the observation of low $x$ resummation effects since higher order QCD corrections lead to 
a good description of data.

\begin{figure}[h]
\centering
\includegraphics[width=4.5in,clip]{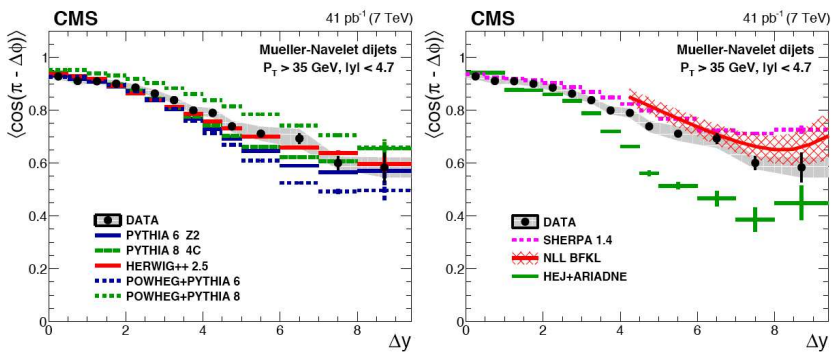}
\caption{Measurement of the azimuthal decorrelation between Mueller Navelet jets measured by the CMS collaboration
and compared to DGLAP Monte Carlo and BFKL calculations.}
\label{mnjets}       
\end{figure}

The second observable that was proposed to look for BFKL resummation effects at hadronic colliders is the Mueller-Navelet 
jet production~\cite{mn}. The idea is to look for jets separated with a large interval in rapidity. When the transverse
momentum of the two jets is similar, the BFKL dynamics predicts a larger cross section than the DGLAP one because
of the $k_T$ ordering of the different gluons along the ladder as before~\cite{mnus,mnwallon}. One possible observable is the measurement of dijet azimuthal decorrelation which is larger for BFKL dynamics because of the multiple
gluon emission~\cite{mnus}. Unfortunately, this 
prediction suffers from high order corrections to BFKL dynamics that has a tendency to bring both predictions to be 
closer to each other. As an example, the most recent measurements by the CMS and ATLAS collaborations~\cite{mnjetcms} are shown in Fig.~\ref{mnjets}. Data can be described by BFKL dynamics as expected but also by NLO DGLAP QCD calculations. 
It is thus needed to define new possible variables or going to larger interval of rapidity between jets, using the CASTOR~\cite{CASTOR}
very forward calorimeter in CMS as an example, in order to enhance further the BFKL contribution.

\begin{figure}[h]
\centering
\includegraphics[width=4.5in,clip]{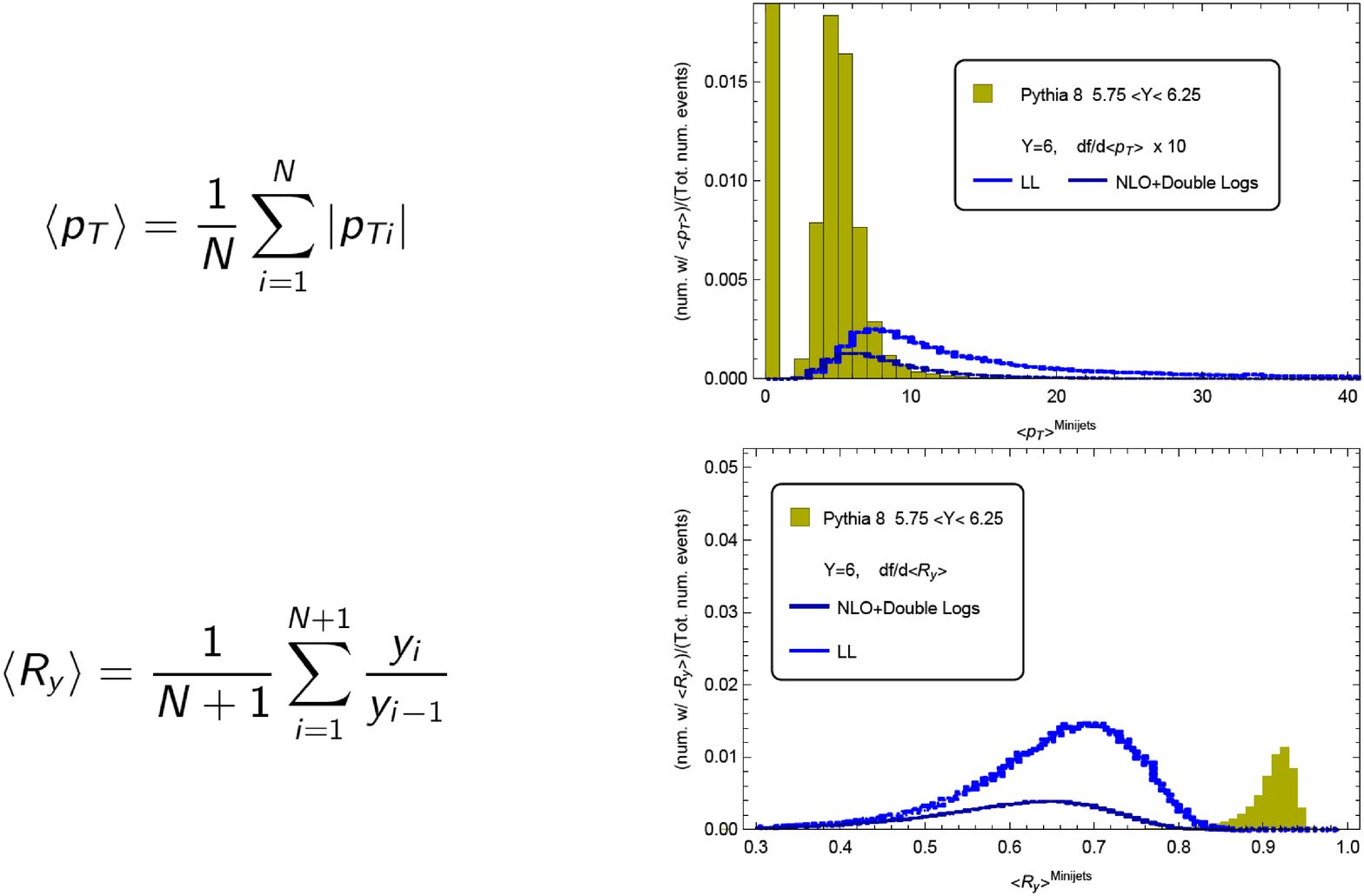}
\caption{New less inclusive measurements in Mueller-Navelet jet events to look for BFKL resummation effects:
average $p_T$  rapidity of mini-jet production.}
\label{mnjdyn}       
\end{figure}

New variables were recently proposed to look for BFKL dynamics~\cite{sabiovera}. The idea is to look for mini-jet production
between the two Mueller-Navelet jets that are predicted by DGLAP or BFKL dynamics as illustrated in Fig.~\ref{mnjdyn}.
The pattern for gluon emission is different between both dynamics because of the $k_T$-ordering of the gluons in the
ladder for the DGLAP prediction. As an example, it would be possible to measure the mini-jet production with transerve momentum above 20 or 25 GeV between the two Mueller Navelet jets with $p_T>$40 or 45 GeV.

In the next section, we will discuss another method to look for BFKL dynamics, namely the measurement of jet gap jet events.

\section{Jet gap jet production}

\subsection{Gap between jets at the Tevatron and the LHC}
The measurement of gap between jets or the Mueller-Tang process~\cite{muellertang} was performed at the Tevatron~\cite{d0jgj} and is also being done at the LHC~\cite{cmsjgj}. The idea is to look for two jets separated by a large interval in rapidity $\Delta \eta$ with a gap devoid of any activity or energy (no particle emission) in a central part of the detector $\Delta \eta_{gap}$
(for instance between (-1) and (1) in rapidity). This means that a colorless object, or a Pomeron, is emitted between the two jets. The natural description of this object is the BFKL Pomeron. The prediction of the BFKL cross section to produce these events is not negligible whereas the DGLAP prediction is very small for gap sizes above two units in rapidity. This is why
this process is ideal to look for BFKL resummation effects. However, it suffers from potential additional interactions, such
as soft gluon exchanges, that can destroy the gap between the jets. This is the so-called the survival probability that can be as 
low as 0.1 at Tevatron center-of-mass energy (about 2 TeV) and 0.03 at the LHC (about 14 TeV).

\subsection{The NLL BFKL formalism and comparison with data}
In order to describe the measurements of gap between jets at the Tevatron and the LHC, the following BFKL formalism was used.
The production cross section of two jets with a gap in rapidity between them reads
\begin{equation}
\frac{d \sigma^{pp\to XJJY}}{dx_1 dx_2 dE_T^2} = {\cal S}f_{eff}(x_1,E_T^2)f_{eff}(x_2,E_T^2)
\frac{d \sigma^{gg\rightarrow gg}}{dE_T^2},
\label{jgj}\end{equation}
where $\sqrt{s}$ is the total energy of the collision,
$E_T$ the transverse momentum of the two jets, $x_1$ and $x_2$ their longitudinal
fraction of momentum with respect to the incident hadrons, $S$ the survival probability,
and $f$ the effective parton density functions. The rapidity gap
between the two jets is $\Delta\eta\!=\!\ln(x_1x_2s/p_T^2).$ 

The cross section is given by
\begin{equation}
\frac{d \sigma^{gg\rightarrow gg}}{dE_T^2}=\frac{1}{16\pi}\left|A(\Delta\eta,E_T^2)\right|^2
\end{equation}
in terms of the $gg\to gg$ scattering amplitude $A(\Delta\eta,p_T^2).$ 

In the following, we consider the high energy limit in which the rapidity gap $\Delta\eta$ is assumed to be very large.
The BFKL framework allows to compute the $gg\to gg$ amplitude in this regime, and the result is 
known up to NLL accuracy (in this calculation, the coupling to the proton, the so-called impact factors, was computed at LL)
\begin{eqnarray}
A(\Delta\eta,E_T^2) 
=  \frac{16N_c\pi\alpha_s^2}{C_FE_T^2}
\sum_{p=-\infty}^\infty \int \frac{d \gamma}{2i \pi}
\frac{[p^2-(\gamma-1/2)^2]}
{[(\gamma-1/2)^2-(p-1/2)^2]} 
\frac{\exp\left\{\bar\alpha(E_T^2)\chi_{eff}[2p,\gamma,\bar\alpha(E_T^2)] \Delta \eta\right\}}
{[(\gamma-1/2)^2-(p+1/2)^2]} 
\label{jgjnll}
\end{eqnarray}
with the complex integral running along the imaginary axis from $1/2\!-\!i\infty$ 
to $1/2\!+\!i\infty,$ and with only even conformal spins contributing to the sum, and 
$\bar{\alpha}=\alpha_S N_C/\pi$ the running coupling. 

Let us first notice that the sum over all conformal spins is absolutely necessary. Considering
only $p=0$ in the sum of Equation~\ref{jgjnll} leads to a wrong normalization and a wrong jet $E_T$
dependence, and the effect is more pronounced as $\Delta \eta$ diminishes. In the following, we 
summed over all conformal spins (practically, we resumed hundreds of terms).

In this study, we performed a parametrised distribution of $d \sigma^{gg\rightarrow gg}/dE_T^2$
so that it can be easily implemented in the HERWIG Monte Carlo~\cite{herwig} since performing the integral over
$\gamma$ in particular would be too much time consuming in a Monte Carlo. The implementation of the
BFKL cross section in a Monte Carlo is absolutely necessary to make a direct comparison with data.
Namely, the measurements are sensitive to the jet size (for instance, experimentally the gap size
is different from the rapidity interval between the jets which is not the case by definition in the
analytic calculation).

The D\O\ Collaboration measured the jet gap jet cross section ratio with respect to the total dijet
cross section, requesting for a gap between -1 and 1 in rapidity, as a function of the second
leading jet $E_T$,
and $\Delta \eta$ between the two leading jets for two different low and high $E_T$ samples
(15$<E_T<$20 GeV and $E_T>$30 GeV). To compare with theory, we compute the following quantity
\begin{eqnarray}
\text{Ratio} = \frac{\text{NLL~HERWIG}}{\text{LO~Herwig}} \times \frac{\text{LO~QCD dijets}}{\text{NLO~QCD dijets}} 
\end{eqnarray}
in order to take into account the NLO corrections on the dijet cross
sections, where \text{BFKL~ NLL HERWIG} and \text{Dijet~HERWIG} denote the BFKL NLL and the dijet cross section
implemented in HERWIG~\cite{herwig}. The NLO QCD cross section was computed using the NLOJet++ program~\cite{nlojet}.

\begin{figure}[h]
\centering
\includegraphics[width=3.in,clip]{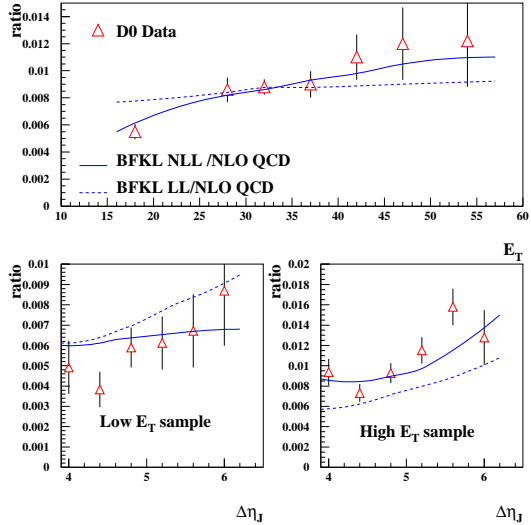}
\caption{Comparison between the measurement of gap between jets events by the DO collaboration and the BFKL calculation as a  function of  jet $E_T$ or $\Delta \eta$ between jets for the low and high $E_T$ samples.}
\label{d0jgj}       
\end{figure}

The comparison with D\O\ data~\cite{d0jgj} is shown in Fig.~\ref{d0jgj}. We find a good agreement between the data
and the BFKL calculation. It is worth noticing that the BFKL NLL calculation leads to a better result
than the BFKL LL one (note that most studies in the literature  considered only the 
$p=0$ component which is not a valid assumption). The comparison with CDF data~\cite{d0jgj} leads to a similar conclusion
and data are well described by the BFKL NLL formalism.

The latest results have been provided by the CMS Collaboration using data collected in pp collisions at $\sqrt{s} = 7$ TeV~\cite{cmsjgj}. Each of the leading two jets has transverse momentum of $p_\text{T} > 40$ GeV, pseudorapidity of $1.4 < |\eta_\text{jet}| < 4.7$, and opposite signed pseudorapidities $\eta_\text{jet 1} \cdot \eta_\text{jet 2} < 0$. The fraction of jet-gap-jet events were measured as a function of $p_\text{T, jet 2}$ and pseudorapidity difference between the leading two jets $\Delta\eta_\text{jj}$.

\subsection{The NLL BFKL formalism including NLO impact factors}
The full NLL BFKL calculation using the NLO impact factor is in progress~\cite{federico1}. 
Several complications appear when the impact factors are refined to NLO.
\begin{itemize}
\item The back-to-back symmetry proper to an elastic interaction is lost since a third parton can appear in the final state.
\item Since the NLO impact factor interference diagrams connect in a non trivial way to the gluon-Green function (GGF)  their convolution cannot be factorized any more into an amplitude square involving only the integrated GGF. Moreover, the GGF in its full form is written in terms of Gauss-hypergeometric functions in some particularly complicated combinations especially for large conformal spins and it complicates  the numerical implementation.
\item A formal divergence appears in the real-corrections published in~\cite{Agustin} when the proper jet reconstruction algorithm is employed. The problematic behavior appears when besides the two tagged jets, a third parton is emitted in the gap region but with a sub-threshold transverse energy. One would think that such configuration should be suppressed when the additional parton becomes  more central. This puzzling behavior motivates an independent cross-check of previously pubished results that is currently a work-in-progress.
\end{itemize}
It is expected that the full NLO calculation will be available soon, leading to more accurate comparison with data.

\subsection{Jet gap jet events in diffraction}

\begin{figure}[h]
\centering
\includegraphics[width=3.in,clip]{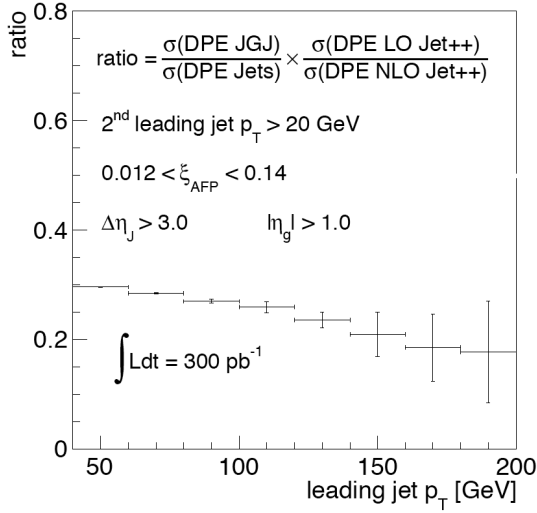}
\caption{Ratio of the jet-gap-jet to the inclusive jet cross sections at the LHC 
as a function of jet $p_T$ in double pomeron exchange events where the protons
are detected in AFP or CT-PPS. }
\label{fig10}       
\end{figure}

In this section, we discuss the observation of gap between jets in  single diffractive or double Pomeron exchange
events~\cite{jgjpap}. These
processes are similar to the one discussed in the previous section but they require in addition the existence of at least one intact proton after interaction.
This process was implemented in the FPMC generator~\cite{FPMC}.

The DPE jet-gap-jet event ratio is defined as the ratio of the cross section for the DPE jet-gap-jet (JGJ) production to the DPE inclusive dijet (Jets) production
$R = \sigma ( DPE\ \text{JGJ} )/\sigma ( DPE\ \text{Jets} )$
and is plotted in Fig.~\ref{fig10} as a function of the transverse momenta of the first-leading jet.
To take into account NLO QCD effects, absent in the FPMC program, the LO 
ratio obtained was corrected by the cross section ratio 
$\sigma ( DPE\ LO\ Jet++ )/\sigma ( DPE\ NLO\ Jet++ )$, obtained by the 
NLO Jet++ program \cite{nlojet}. 

As far as the gap fraction $\sigma ( DPE\ JGJ )/\sigma ( DPE\ Jets )$ is 
concerned, note that we did not consider an additional suppression factor 
for DPE jet-gap-jet production, on top of the 0.03 of DPE inclusive 
jet production. Therefore, in the predictions of Fig.~\ref{fig10}, all the rapidity gap survival probability cancel. We would like to point out that this last point is an assumption, DPE jet-gap-jet production could be subject to a bigger suppression than DPE inclusive jet production, due to the extra color-singlet exchange in the hard cross section. However, we do not expect this potential additional factor (on top of 0.03) to be large, due to the fact that extra soft interactions with the BFKL pomeron are unlikely: the 2-to-2 hard scattering takes place on much shorter time scale compared to the soft interactions filling the rapidity gaps. In any case, this will be checked at the LHC and if necessary, our numbers can then be adjusted accordingly.

\section{Conclusion}
In this article, we described different methods to look for  BFKL resummation effects at the LHC. Muller Navelet jets are traditionally used 
as a BFKL like signature but they do not lead to a clear distinction between DGLAP and BFKL resummation effects. The possibility is to study
more exclusive variables (such as the mini-jet production between the two Mueller Navelet jets) or to look for gap between jets in hadron-hadron interactions.
We computed the BFKL NLL jet gap jet cross section (including LO impact factors) that leads to a good description of Tevatron data.
The full NLO BFKL cross section calculation  including NLO impact factors is in progress and should be released soon.

\end{document}